\newcommand{\CLASSINPUTtoptextmargin}{0.78in}
\documentclass[conference]{IEEEtran}
\IEEEoverridecommandlockouts

\usepackage{array}
\usepackage{amsmath,amssymb,amsfonts}
\usepackage{algorithmic}
\usepackage{graphicx}
\usepackage{textcomp}
\usepackage{xcolor}
\usepackage[caption=false]{subfig}
\usepackage{url}
\def\BibTeX{{\rm B\kern-.05em{\sc i\kern-.025em b}\kern-.08em
    T\kern-.1667em\lower.7ex\hbox{E}\kern-.125emX}}
\begin{document}

\title{How to make Firmware Updates over LoRaWAN Possible\\
}

\author{\IEEEauthorblockN{Khaled Abdelfadeel}
\IEEEauthorblockA{\textit{School of Computer Science and IT} \\
\textit{University College Cork}\\
Cork, Ireland \\
khaled.abdelfadeel@ieee.org}
\and
\IEEEauthorblockN{Tom Farrell and David McDonald}
\IEEEauthorblockA{\textit{Danalto} \\
Dublin, Ireland \\
first.last@danalto.com}
\and
\IEEEauthorblockN{Dirk Pesch}
\IEEEauthorblockA{\textit{School of Computer Science and IT} \\
\textit{University College Cork}\\
Cork, Ireland \\
d.pesch@cs.ucc.ie}
}

\maketitle

\begin{abstract}

Embedded software management requirements due to concerns about security vulnerabilities or for feature updates in the Internet of Things (IoT) deployments have raised the need for Firmware Update Over The Air (FUOTA). With FUOTA's support, security updates, new functionalities, and optimization patches can be deployed with little human intervention to embedded devices over their lifetime. However, supporting FUTOA over one of the most promising IoT networking technologies, LoRaWAN, is not a straightforward task due to LoRaWAN's limitations that do not provide for data bulk transfer such as a firmware image. Therefore, the LoRa Alliance has proposed new specifications to support multicast, fragmentation, and clock synchronization, which are essential features to enable efficient FUOTA in LoRaWAN. In this paper, we review these new specifications and evaluate the FUOTA process in order to quantify the impact of the different FUOTA parameters in terms of the firmware update time, the device's energy consumption, and the firmware update efficiency, showing different trade-offs among the parameters. For this, we developed FUOTASim, a simulation tool that allows us to determine the best FUOTA parameters.

\end{abstract}


\IEEEpeerreviewmaketitle

\section{Introduction}

Firmware Update Over The Air (FUOTA) defines the process of updating a device's firmware over a wireless medium. This is a crucial feature for large-scale wireless sensor network installations as it allows to deploy security, optimization, and/or new-functionality patches without much human intervention in order to protect devices, extend their lifetime and/or enhance their performance ~\cite{enablingfuota}. In LoRaWAN~\cite{lorawan2017specs}, FUOTA is even a more critical requirement because of the long device lifetime that LoRaWAN promises, e.g., 10 years~\cite{appsurvey}. This long lifetime stands in contradiction to the fast-changing modern software life-cycle and the LoRaWAN standard, which is subject to continued development. Therefore, FUOTA represents a way to keep LoRaWAN devices up-to-date with the standard throughout their lifetime, which important for reliable, safe and secure long-term operation.

The nature of FUOTA requires downloading a big block of data (e.g. a few hundreds of kilobytes) to the devices. This is a challenging task in LoRaWAN because of the limitations of the communications technology itself~\cite{adelantado2017understanding}. LoRaWAN is low data-rate technology, offering at most a few 10s of kbits/s. LoRaWAN also operates in the unlicensed sub-GHz band, where transmissions have to obey a duty cycle restriction, for instance, 1\% in Europe. In this case, a device/gateway has to be silent for at least 99 times the last transmission time after each transmission. In addition to that, LoRaWAN is designed for applications with predominantly uplink transmissions. For instance, a downlink transmission, in case of a class-A device, is only available after an uplink transmission. All these limitations challenge efficient FUOTA over LoRaWAN.

In order to better understanding the impact of LoRaWAN limitations on FUTOA, let us consider how to transmit a firmware image of 50 kBytes using DR2 (SF10/ 125 kHz). Even if the maximum packet size (i.e. 51 bytes for DR2) is used, about 1004 downlink packets are required to transmit the whole firmware. For this, a similar number of uplink transmissions is required to solicit the downlink transmissions. Even with a perfect wireless channel with no losses, the firmware update would take ca. 17 hours to upgrade only one device because of the 1\% duty cycle limitation. Consequently, updating a medium-size deployment could take up to a few weeks, which is not practical.

The above example points to a number of features that are required for LoRaWAN in order to support efficient FUOTA:
\begin{itemize}
    \item A mechanism to send downlink transmissions without the need for uplink transmissions to be sent first. This optimizes the devices' duty cycle and, thus, their power consumption.
    \item Multicast support in order to optimize the gateways' duty cycle by sending downlink transmissions to multiple devices simultaneously.
    \item A mechanism to download a big data block and recover packet losses without congesting the medium with transmissions to request the missing packets.
\end{itemize}
For this, the LoRa Alliance, the industry body behind the LoRaWAN standard, created the FUOTA working group to define the baseline needs to enable efficient FUOTA over LoRaWAN. This has resulted in new specifications to cover multicast, fragmentation and time synchronization topics, which are essential features for efficient FUOTA. 

In this paper, we describe these new LoRaWAN specifications and examine how the new features can enable fast and efficient firmware update. Additionally, we analyze the proposed FUOTA process in order to quantify the impact of the different parameters and show the trade-offs among them. In order to analyse the process, we developed a new simulation tool, FUOTASim, to study the FUOTA process. FUOTASim can support LoRaWAN operators to determine the best parameters when performing FUOTA. To the best of our knowledge, this work is the first scientific paper that considers FUTOA over LoRaWAN, which is one of the most challenging networks for over the air software updates.

\section{Preliminaries}
This section highlights some of the LoRaWAN features in addition to the new specifications, which have been developed to support efficient FUOTA.

\begin{table}[!tb]
  \centering
  \caption{LoRaWAN Regional Parameters in Europe} \label{tab:lorawan_europe}
  \begin{tabular} {l l l || l l}
  \hline
  Data & Configu- & Max App & Default        & Duty \\
  Rate & rations  & Payload & Channels & Cycle\\
  \hline
  0 & SF12/125KHz & 51 bytes&             &   \\
  1 & SF11/125KHz & 51 bytes& 868.10MHz (U/D)& 1\% \\
  2 & SF10/125KHz & 51 bytes& 868.30MHz (U/D)& 1\% \\
  3 & SF9/125KHz & 115 bytes& 868.50MHz (U/D)& 1\% \\
  4 & SF8/125KHz & 222 bytes& 869.525MHz (D)& 10\%\\
  5 & SF7/125KHz & 222 bytes&             &   \\
  \hline
  \end{tabular}
\end{table}

\subsection{LoRaWAN}

LoRaWAN~\cite{lorawan2017specs,adelantado2017understanding,appsurvey} defines the Medium Access Control (MAC) rules, the system architecture, and regional parameters for operation in different regions of the world. In this work, we consider the regional parameters for Europe as shown in Table~\ref{tab:lorawan_europe}. LoRaWAN supports six data rates (i.e. configured by the spreading factor and the bandwidth), thanks to its unique modulation, called LoRa~\cite{semtech2015lora}. LoRaWAN operates in the EU863-870 ISM band, and by default, three channels (868.10, 868.30, and 868.50 MHz) with 1\% duty cycle are supported for uplink and downlink transmissions in addition to one channel (869.252 MHz) with 10\% duty cycle for downlink transmissions.

\begin{figure}[!tb]
  \centering
  \includegraphics[width=1\columnwidth]{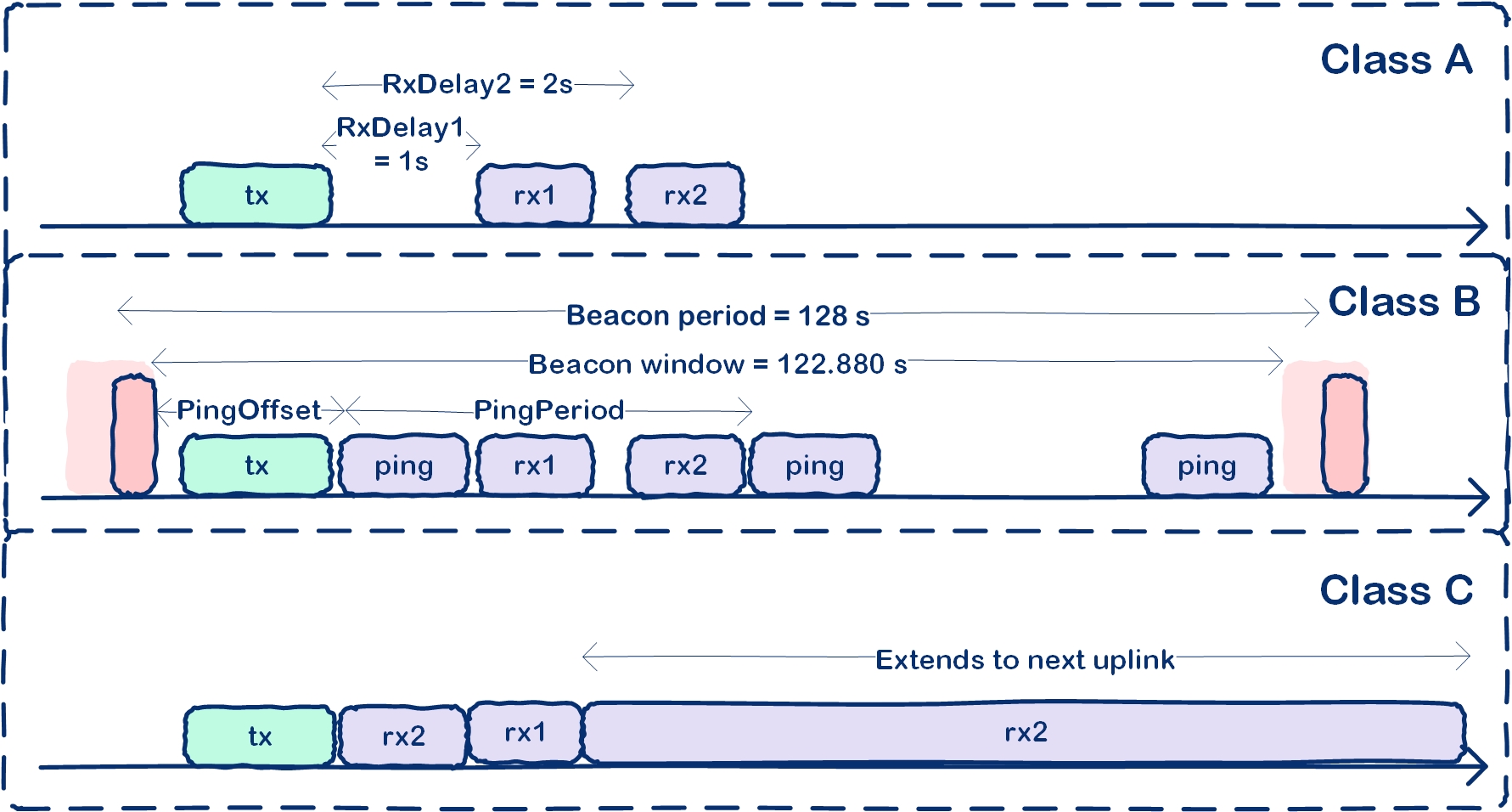}
  \caption{LoRaWAN Classes of operations}\label{fig:lorawan_classes}
  \vspace{-0.35cm}
\end{figure}

LoRaWAN supports three classes of operation, namely A, B, and C, where each class offers different downlink capabilities to suit a range of IoT applications as shown in Fig.~\ref{fig:lorawan_classes}. Uplink transmissions in LoRaWAN follow a simple ALOHA protocol, where devices transmit whenever they have data to transmit as long as the duty cycle permits it and without performing any sort of listen-before-talk policy.

Class A is the mandatory profile that all LoRaWAN devices have to support. In this class, each uplink transmission is followed by two receive windows at specific times. Downlink transmissions are only allowed at the beginning of these receive windows. The downlink transmission in the first window is performed using the same configurations (i.e. data rate and channel) as the previous uplink transmission. However, a fixed configurations, i.e., DR0 (SF12/125KHz) on 869.525 MHz is used in the second window. On the contrary, class C permits downlink transmissions all the time except when the devices transmit. This is done by extending the second receive window until the next uplink transmission, resulting in huge power consumption as devices remain in a receive mode for most of the time. 

Class B allows more receive windows than class A but without the huge power consumption of class C. Besides the two regular receive windows after each uplink transmission, extra periodic receive windows, called ping slots, are opened at synchronized times. The synchronization is guaranteed by receiving the gateway beacons that are sent periodically every 128 secs. 
The usable time period between two beacons is called beacon window and it is divided into $2^{12}=4096$ ping slots of 30 ms each numbered from 0 to 4095. The ping slot periodicity, \textit{pingPeriod}, of a device is defined using $0.96\times2^{p}$ secs, where $0<=p<=7$. For a certain periodicity $p$, the assigned number of ping slots in a beacon window is defined using $2^{7-p}$. In case of $p=0$, a device opens 128 ping slots, one slot almost every 1 sec. In case of $p=7$, only one ping slot is opened every 128 secs, which is the maximum supported ping slot period. In order to avoid systematic collisions, \textit{pingOffset} is calculated at the beginning of each beacon period to indicate the time of the first ping slot. \textit{pingOffset} is a randomised offset, whose values can range from 0 to ($2^{7-p}-1$).

\subsection{Key Requirements of FUOTA}

The FUOTA working group defined the baseline needs to perform FUOTA over LoRaWAN. These needs have been described in new specifications that we highlight here. The objective of the new specifications (mutlicast, fragmentation, and clock synchronization) is to standardize this essential process, leading to an interoperable FUOTA solution. It is worth mentioning that the new specifications are not part of the LoRaWAN MAC but run at the application layer.

\begin{figure*}[!tb]
  \centering
  \vspace{0.35cm}
  \includegraphics[width=1.7\columnwidth]{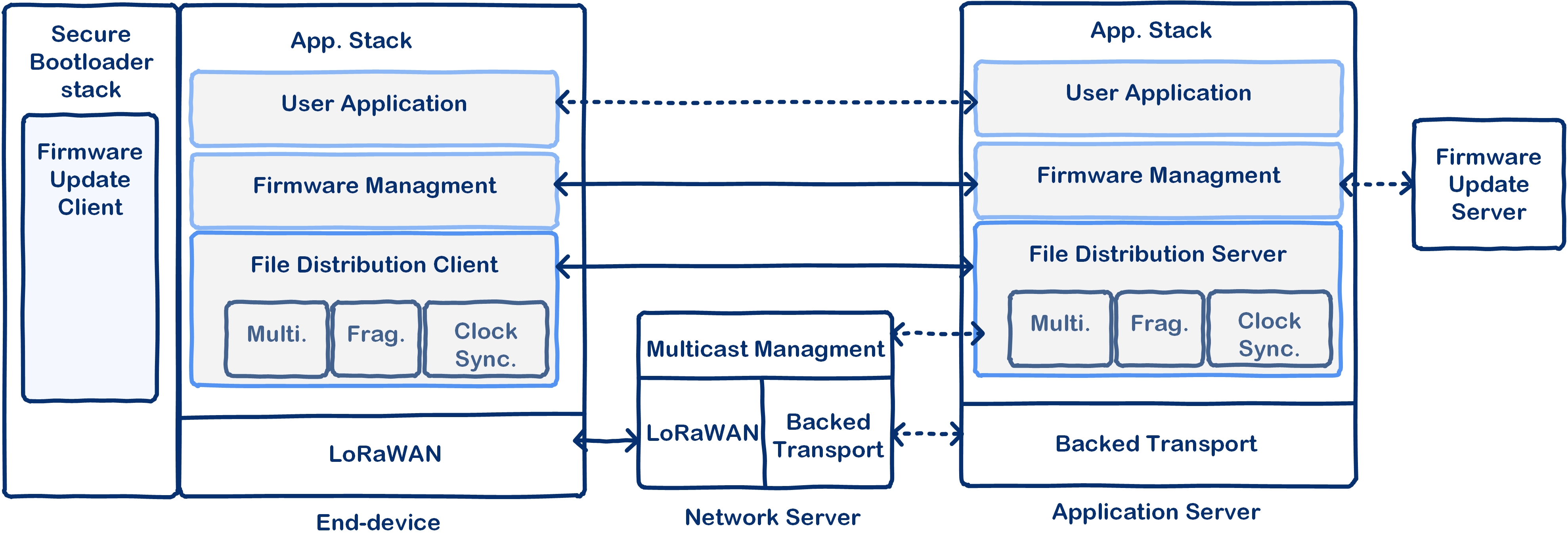}
  \caption{FUOTA Architecture~\cite{fuotaprocesssummary}}\label{fig:fuota_arch}
  \vspace{-0.35cm}
\end{figure*}

Fig.~\ref{fig:fuota_arch} shows the recommended FUOTA architecture~\cite{fuotaprocesssummary}, where interfaces with solid lines are described in the LoRa Alliance specifications, otherwise, they are out of the LoRa specifications scope. The architecture shows the network server in the middle, which handles the communication between LoRaWAN devices and application servers. In addition to that, the network sever manages the multicast, including creation, deletion and/or editing of multicast groups and assures delivery of multicast downlink transmissions to all devices in a group. This is performed by calculating the minimum set of gateways that have to send the same multicast transmission to cover all devices in a group. Firmware update server (right-hand side) together with the firmware update management initiate and run the FUOTA process for a list of devices. These two components are the brain of the FUOTA process that generates the firmware image and controls the process and the file distribution server. The file distribution server is an application that is tasked to deliver the firmware image to a group of devices, using the underlying features, i.e., multicast, fragmentation, and clock synchronization. These features are highlighted later in this subsection.

At the device side, the counterpart components of the firmware management and the file distribution server are present in addition to an additional stack for a secure bootloader, which is essential for any device performing firmware updates. The bootloader is responsible for checking the availability and the integrity of the new firmware image and overwriting the old firmware with the new one. This part of the process is out-of-scope of this paper. However, we would like to refer to the work-in-progress of the Software Update for IoT (SUIT) working group\footnote{\url{https://datatracker.ietf.org/wg/suit/about/}}. SUIT is chartered by the Internet Engineering Task Force (IETF) to standardize a manifest that provides meta-data about the firmware image (such as a firmware identifier, the hardware the firmware needs to run, and dependencies on other firmwares), as well as cryptographic information for protecting the firmware image in an end-to-end fashion. The new solutions are mainly targeting constrained IoT devices similar to LoRaWAN devices~\cite{zandberg2019secure}.

\subsubsection{Multicast~\cite{multicastspec}}\label{multicast} the objective is to have a group of class A devices receive the same downlink transmission at the same time. This requires that the group of devices is in a receive mode at the same time and share the same security keys to be able to decrypt the same downlink transmission. For this, the multicast specification defines a command to program a receive distribution window of class C or class B into a group of class A devices. Additionally, the specification defines commands to instruct the group of devices to switch to class C or B temporarily at the beginning of the receive window and switch back to their normal class at the end of the receive window. All commands of this specification are sent to each device individually using unicast messages and 200 as a port number.

Multicast command \textit{McGroupSetupReq} is sent to a device to set up a multicast group. The multicast security key is sent in this command to be used by all devices in the group to derive the multicast security application and network keys. These keys are used to encrypt the multicast messages so by sharing the same keys, the group can decrypt the same multicast messages. \textit{McGroupSetupAns} is sent back by a device to acknowledge the multicast setup. 
Subsequently, command \textit{McClassCSessionReq} is sent to a device to indicate that the group that has been set up earlier is a class C group. In addition to that, the command defines the session time, the session time out, the data rate and the channel. The session time indicates when the device has to start the class C receive window. The time is expressed as the time in seconds since 00:00:00, Sunday 6\textsuperscript{th} of January 1980 (start of the GPS epoch) modulo $2^{32}$. The session time out indicates the maximum length in seconds for a device to stay in class C before reverting to class A. The data rate and the channel are the configurations that will be used to send the multicast transmissions. For a class B multicast group, command \textit{McClassBSessionReq} is sent, which is similar to the \textit{McClassCSessionReq} command, except in this command, the slot ping periodicity of class B is defined, which indicates how many ping slots are assigned and their periodicity for multicast transmissions. For acknowledgment, \textit{McClassCSessionAns} and \textit{McClassBSessionAns} are sent back by a device to acknowledge \textit{McClassASessionReq} and \textit{McClassBSessionReq}, respectively.
 
\subsubsection{Clock Synchronisation~\cite{clocksyncspec}} As LoRaWAN devices usually do not have access to accurate clocks, their times are not reliable for performing \textit{McClassCSessionReq} and \textit{McClassBSessionReq} commands. Therefore, the clock synchronization specification defines a way for the devices to correct their clock skews. The basic idea is that the network has access to an accurate GPS clock that can be used to correct the devices' clocks. All commands of this specification are sent as application messages and 202 port number is used to distinguish the application. Command \textit{AppTimeReq} is sent by a device to ask for a clock correction. The command includes the device time, which indicates the current device clock. The time is again expressed as the time in seconds since 00:00:00, Sunday 6\textsuperscript{th} of January 1980 (start of the GPS epoch) modulo $2^{32}$. Next, the device gets \textit{AppTimeAns} back, including the time correction that stipulates the time delta correction in secs. The expected accuracy of this approach is around one second, which is enough to run the multicast commands efficiently .

\subsubsection{Fragmentation~\cite{fragmentationspec}} A firmware image is usually quite big (i.e. a few hundreds kBytes), which cannot fit in one downlink packet but needs quite a number of packets. In addition to that, LoRaWAN links are lossy and there is no way to know which packets are lost at which devices during multicast transmissions. Therefore, a mechanism to handle big data blocks and to recover packet losses in a scalable manner is required. For this, the fragmentation specification supports all necessary commands to transport a large data block to one device or to a group of devices reliably if multicast class C or class B is used. All commands of this specification are sent as application messages and port number 201 is used to distinguish this application. 

Command \textit{FragSessionSetupReq} is sent to a device to define a fragmentation session. The command specifies which multicast groups are allowed as input for this fragmentation session. In addition to the number of fragments, the fragment size, the fragmentation algorithm (recovery algorithm), and the padding size are specified. The padding size is used as the firmware image may not be an integer multiple of the fragmentation size. A device sends \textit{FragSessionSetupAns} back to acknowledge setting up the fragmentation session.

At the time of writing of this paper, only one fragmentation algorithm was defined. The algorithm proposes adding a simple forward error correction code to the original firmware image before sending it. This allows devices to \textit{autonomously} recover a certain ratio (based on the code used) of the lost transmissions without requesting re-transmission of lost fragments. This is done by, first, chunking the original firmware image to fragments equal in size and then adding redundancy fragments, which are XORed to some of the original fragments. Devices can use redundant fragments to reconstruct their missing fragments. For example, 5\% redundancy added to the original firmware image allows devices to loose roughly 5\% of the incoming transmissions and still be able to reconstruct the original firmware. 

\begin{figure}[!tb]
  \centering
  \includegraphics[width=1\columnwidth]{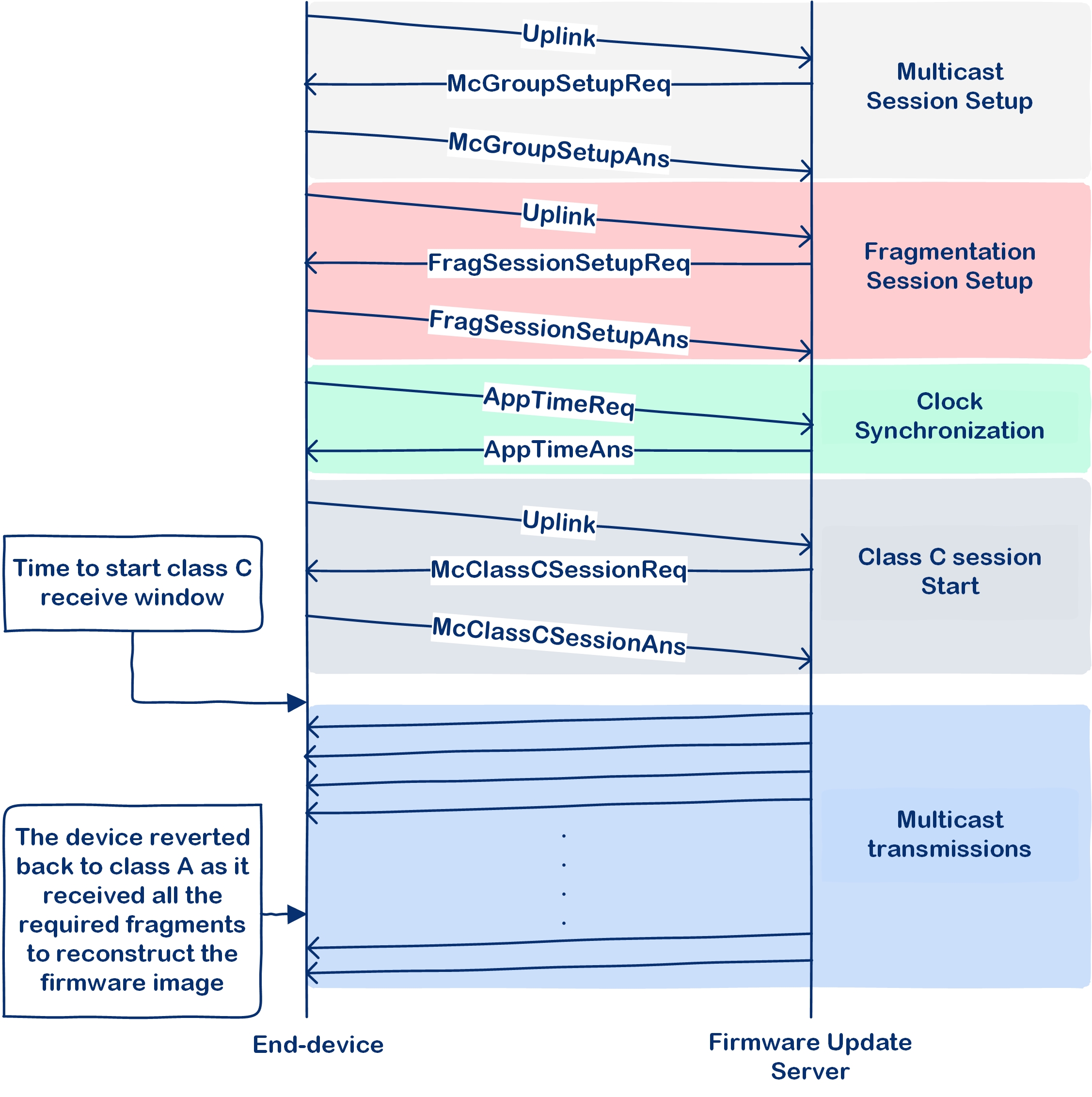}
  \caption{FUOTA Session using class C multicast}\label{fig:fuota_session}
  \vspace{-0.35cm}
\end{figure}

\section{FUOTA Process} \label{fuota_ruotine}

The FUOTA process is initiated at the firmware update server, which generates the new firmware fragments along with the redundant fragments. Also, together with the firmware management, the firmware update server assures that the required sessions (mutlicast, fragmentation and etc.) are already established at the intended devices before sending the fragments. The firmware update server takes system decisions which affect the efficiency of the process. These decisions include the topology of the multicast group(s), the class of the multicast (i.e. C or B), the data rate and the channel to be used to transmit the multicast fragments. For example, the firmware update server may decide to divide a big group of devices into two smaller groups and run two FUOTA sessions in parallel instead of running one FUOTA session. The questions that arises is whether this decision would make the update any more efficient?

Although the LoRa Alliance tries to standardize the FUOTA process by defining the new specifications, the FUOTA routine itself is not standardized and open for contributions. In this section, we present a straightforward FUOTA routine that does not convey any kind of smartness. Nevertheless, the proposed FUOTA routine can help us to study and evaluate the process and define the trade-offs in the system design. This can help LoRaWAN operators understand the impact of their system decisions on the process' efficiency and, thus, can help them to devise smarter FUOTA processes.

Let us consider a firmware update is scheduled for a LoRaWAN deployment, consisting of class A devices. Consequently, the firmware update server configures the same multicast and fragmentation sessions for all devices. An example of the FUOTA session that uses mutlicast class C is shown in Fig.~\ref{fig:fuota_session}. It should be noted that the multicast and the fragmentation requests are downlink commands and, thus, uplink messages are needed first in order for the devices to open receive windows. Once these two sessions are set up, every device sends \textit{AppTimeReq} command to ask for clock correction before the firmware update server can set up the start time of the multicast transmissions. The start time has to be sufficiently far in the future to guarantee that all devices have set up the required sessions before sending the multicast transmissions. For the multicast class B group, the corresponding commands are used as shown in~\S\ref{multicast}. 

At the exact declared time, all devices must switch to class C and open a continued receive window based on the configurations (data rate and channel) that have been sent in the \textit{McClassCSessionReq} command. At the same time also, the firmware update server schedules the firmware fragments one after another until all the fragments, including the redundant ones, have been sent. Once a device receives enough fragments to reconstruct the firmware image, it reverts back to class A. It should be noted that all transmissions, either uplink or downlink are governed by the duty cycle limitations of the channel used. Following the complete download, an integrity check is done on the received image (details are out-of-scope) and the image is marked as ready if the check is passed. Next, the old firmware is replaced with the new one (details are out-of-scope), which completes the firmware update.


\section{Performance Evaluation}

To evaluate the FUTOA process, we developed a simulation tool, called FUOTASim, which leverages the Simpy package for process-based discrete-event simulation in Python. FUOTASim implements the dual-component log-distance pathloss model from~\cite{abdelfadeel2019surveys}, which was fitted to real LoRaWAN measurements. FUOTASim also considers the packet error model that was presented in~\cite{abdelfadeel2019free}, which draws on a probabilistic reception model based on the signal strength and packet length. In addition to that, FUOTASim adopts some features from FREESim~\cite{abdelfadeel2019free} such as the impact of the imperfect orthogonality of spreading factors and the duty cycle limitations, leading to realistic simulation results. Finally, FUOTASim simulates the FUOTA process as described in \S\ref{fuota_ruotine} with varying parameter settings. The settings allow choosing between multicast class C or class B to perform FUOTA. In the case of class B, a parameter to configure the ping slot periodicity is presented. In addition to that, FUOTASim allows the use of different data rates, firmware sizes, fragment sizes, and redundant codes to perform FUOTA. The aforementioned features are required for a proper evaluation of the FUOTA process, making FUOTASim a useful tool for the LoRaWAN community\footnote{\url{https://github.com/kqorany/FUOTASim}}.

\begin{table}[!tb]
  \centering
  \caption{Simulation Parameters} \label{tab:simparameters}
\begin{tabular}{l|l}
	\textbf{Parameters} & \textbf{Value [Unit] + Comment} \\
    \hline    
    \hline
    Random Seeds & 10  \\
    Devices & 100 - 500 \\
    Rx1 Window & same configurations as previous uplink\\
    Rx2 Window & DR0 and 869.252 MHz (10\% DC)\\
    Data Rate Distribution & [DR0$\rightarrow$6\%, DR1$\rightarrow$8\%, DR2$\rightarrow$8\%, \\
    & DR3$\rightarrow$11\%, DR4$\rightarrow$22\%, DR5$\rightarrow$45\%] \\
    Gateway Receptions & 8 in parallel\\
    LoRaWAN MAC Header & 8  [Bytes] \\
    Path Loss \cite{abdelfadeel2019surveys} & \underline{near} ($< 400m, d_{0}=92.67,$ \\
                        & $PL_{d_0}=128.63, \gamma=1.05, \sigma=8.72$) \\ 
    & \underline{far} ($\ge 400m, d_{0}=37.27,$ \\
    & $PL_{d_0}=132.54, \gamma=0.8, \sigma=3.34$) \\
    Devices Antenna Gain & 2.2 dBm \\
    Gateways Antenna Gain & 8 dBm \\
    Application Uplinks & 15 [Bytes] \\
    Multicast Transmissions & 869.252 MHz (10\% DC) \\
    Redundant Fragments & 30 \\
    Capacity of Batteries & 1000 [mAh], 11100 [Joules]\\
    Power Consumption   & 132 [mW] (Transmission)\\
                        & 48  [mW] (Reception)\\
                        & 0.018   [mW] (Ideal)\\
    \end{tabular}
\end{table}

The simulation campaigns consider one gateway that is placed in the middle of a LoRaWAN deployment. Class A LoRaWAN devices are spatially distributed around that gateway in such a way so as to acquire a certain data rate distribution across the deployment. We aim for a data rate distribution such that 45\% of the devices use DR5 (SF7/125KHz), 22\% use DR4 (SF8/125KHz), 11\% use DR3 (SF9/125KHz), 8\% use DR2 (SF10/125KHz), 8\% use DR1 (SF11/125KHz), and finally 6\% use DR0 (SF12/125KHz). This distribution is obtained from a real LoRaWAN deployment in Dublin, Ireland, making our results more realistic. Table~\ref{tab:simparameters} summarizes all the simulation parameters used in the evaluations. Each simulated study is executed 10 times using different random seeds and the mean across all results is presented along with the standard deviation.

The simulations are divided into two phases: \textit{initial} and \textit{multicast} to show the impact of each phase separately. The initial phase covers everything required before the multicast transmissions can be sent. From Fig.~\ref{fig:fuota_session}, the initial phase covers the multicast session setup, the fragmentation session setup, the clock synchronization, and the class C session start. However, the multicast phase covers the multicast transmissions. The evaluation results are presented in terms of the following metrics:
\begin{itemize}
    \item \textit{energy consumption}, which indicates the average device's energy consumption.
    \item \textit{time}, which indicates the average time required to finish a certain task.
    \item \textit{update efficiency}, which indicates the average ratio of devices that receive the firmware image successfully.
\end{itemize}

\begin{figure}[!tb]
  \centering
    \includegraphics[width=1\columnwidth]{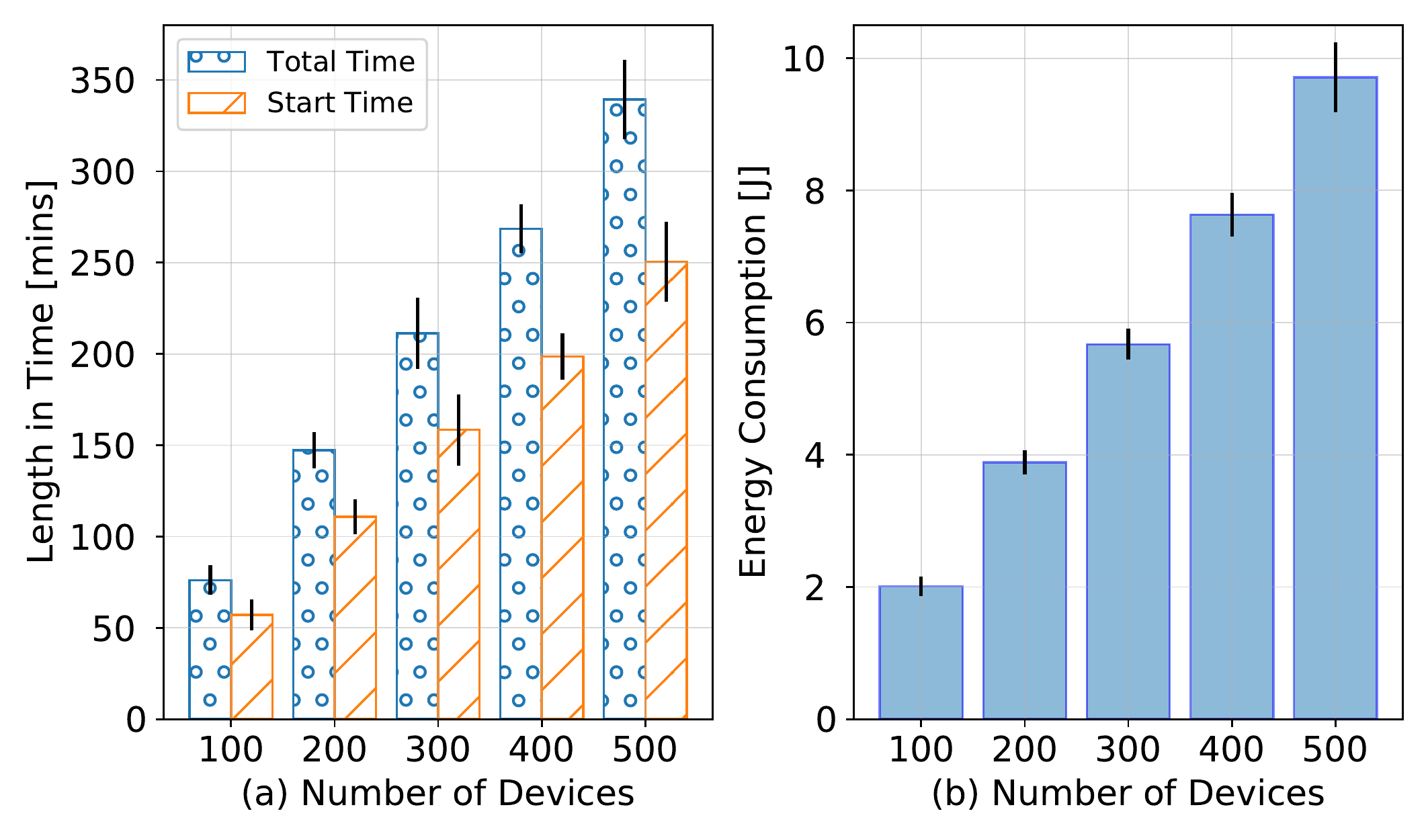}
  \caption{Initial Phase - Time Required and Energy Consumption}\label{fig:timeenergy}
  \vspace{-0.35cm}
\end{figure}

\subsection{Initial Phase Study}

In this phase, we investigate the device's energy consumption (\textit{see} Fig.~\ref{fig:timeenergy}b) and the total time required (\textit{see} Fig.~\ref{fig:timeenergy}a (Total Time)) for the devices to go through the initial phase. Fig.~\ref{fig:timeenergy}b shows a linear increase in the device's energy consumption over the network size, where the energy consumption increases by approx. 2 Joules with every 100 devices added to the network. A similar trend is observed between the total time and the network size (\textit{see} Fig.~\ref{fig:timeenergy}a (Total Time)). Fig.~\ref{fig:timeenergy}a presents also the \textit{start time} metric, which indicates the minimum time required for the devices to complete only the \textit{class C session start} as in Fig.~\ref{fig:fuota_session}. The start time in \textit{McClassCSessionReq} commands has to be sufficiently far in the future to guarantee that every device receives a \textit{McClassCSessionReq} command and acknowledges the command's receipt. In Fig.~\ref{fig:timeenergy}a, the start time metric indicates the \textit{earliest time} for the firmware update server to start the multicast transmissions. If the start time is set to be less than the shown results, some devices may miss their \textit{McClassCSessionReq} commands and, thus, miss the multicast transmissions. The metric is shown in minutes, where the reference time is the time of sending the first \textit{McClassCSessionReq} command ever during the initial phase. The linear increase is also observed here between the start time metric and the network size.

\begin{figure}[!tb]
  \centering
    \includegraphics[width=1\columnwidth]{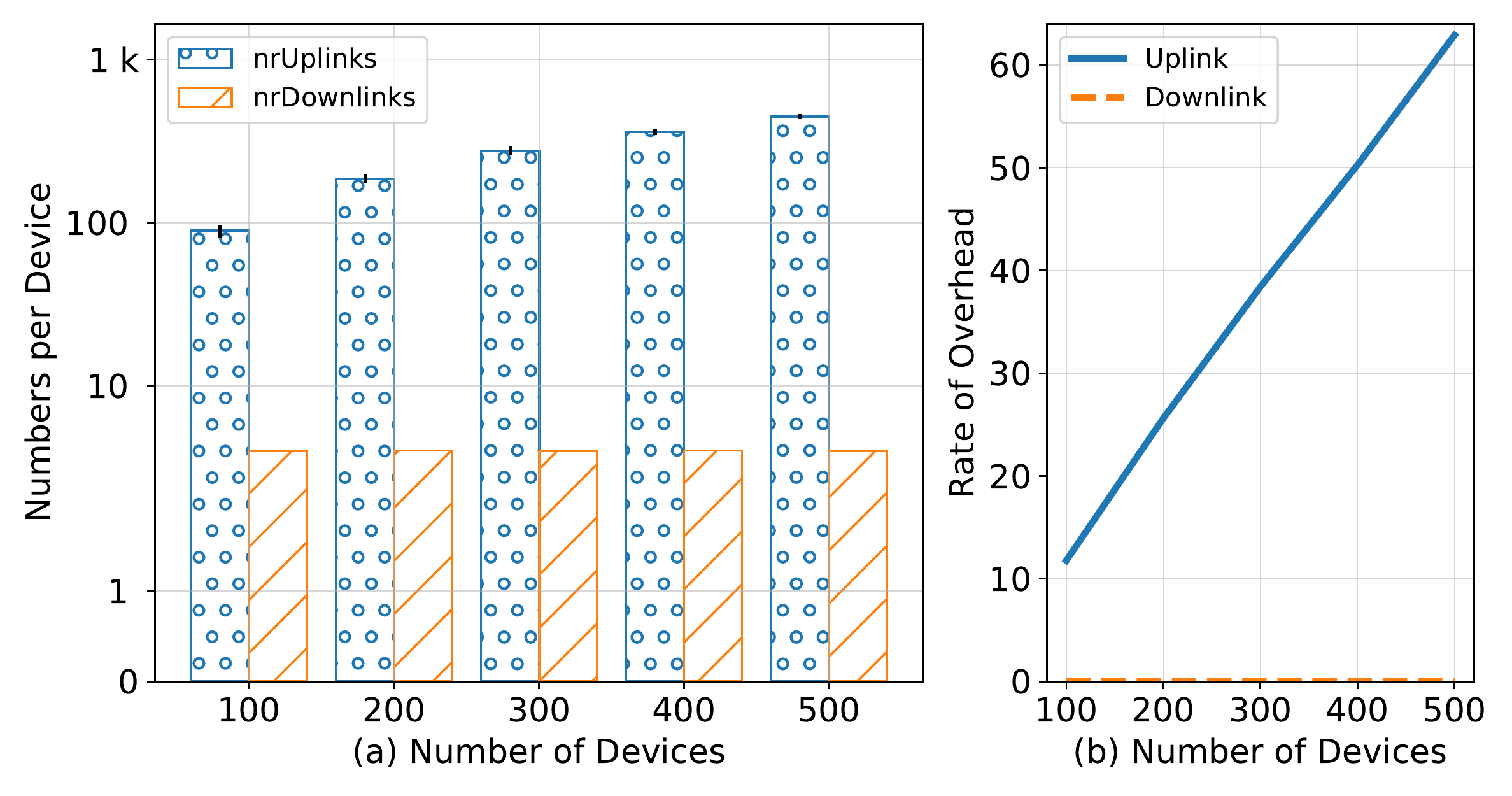}
  \caption{Initial Phase - Uplinks vs Downlinks}\label{fig:upsvsdowns}
  \vspace{-0.35cm}
\end{figure}

\begin{figure}[!tb]
  \centering
    \includegraphics[width=1\columnwidth]{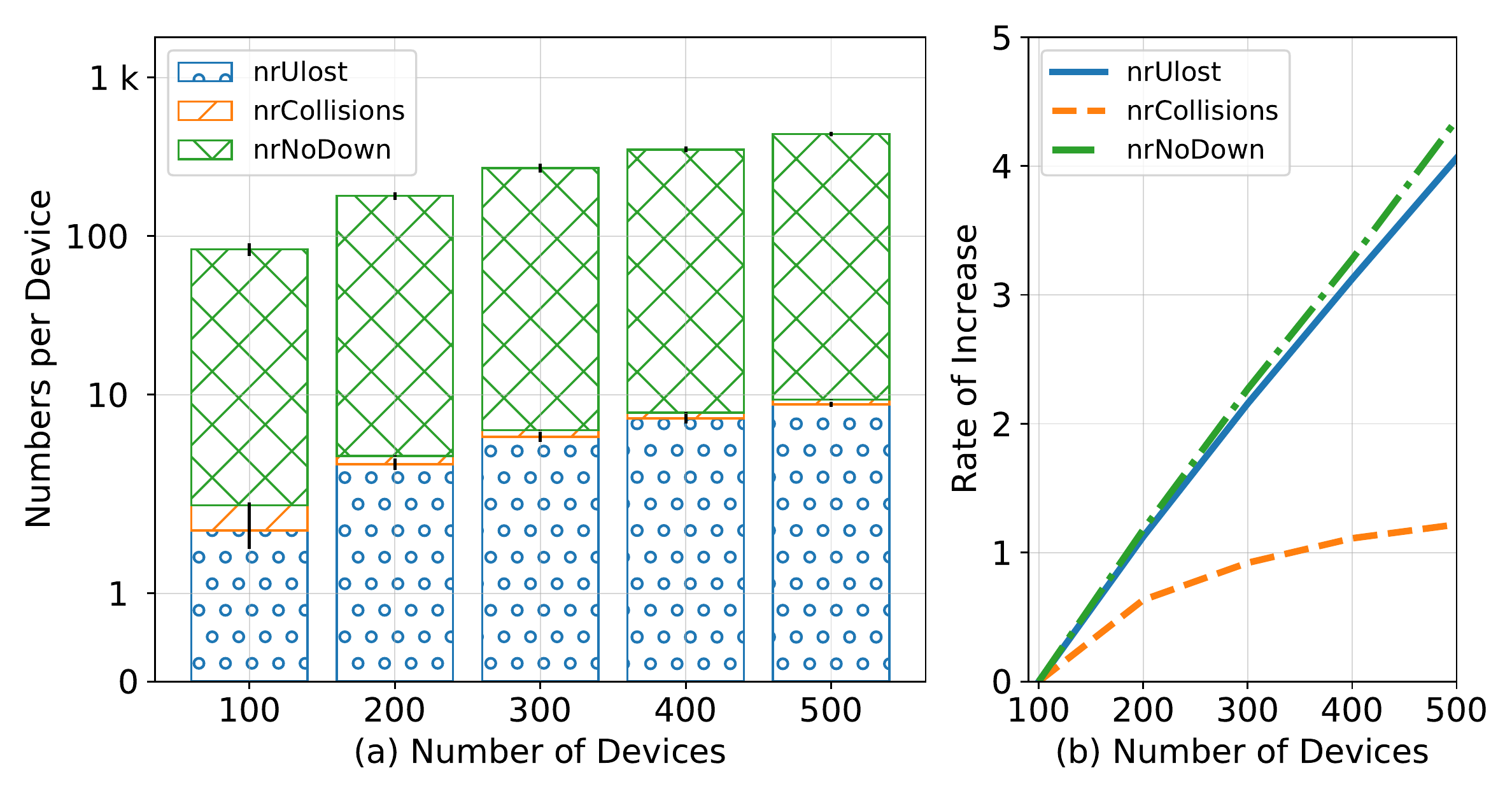}
  \caption{Initial Phase - Source of Losses in Uplink Transmissions}\label{fig:sourceoflosses}
  \vspace{-0.35cm}
\end{figure}

End-devices throughout the initial phase rely on the LoRaWAN MAC (i.e. simple Aloha), which is known for its poor scalability, which is even worse when downlink transmissions are required~\cite{pop2017does}. This is the main reason behind the linear increasing trend we observed in the energy and the time metrics (\textit{see} Fig.~\ref{fig:timeenergy}). In order to quantify the impact of the scalability issue, Fig.~\ref{fig:upsvsdowns}a shows the average number of uplink and downlink transmissions per device throughout the initial phase. In the case of no losses and no duty cycle limitations, only 7 uplink and 4 downlink transmissions are required for each device (\textit{see} Fig.~\ref{fig:fuota_session}). However, in real conditions, the number of uplink transmissions increases significantly (\textit{see} Fig.~\ref{fig:upsvsdowns}a). For instance, in a network with 100 devices, a device sends approx. 94.7 uplink transmissions, which equals approx. 12.5 times more overhead (\textit{see} Fig.~\ref{fig:upsvsdowns}b). With increasing the network size, the overhead increases linearly for the uplink transmissions (\textit{see} Fig.~\ref{fig:upsvsdowns}b). For the downlink transmissions, the losses are very low because of no collisions and the high antenna gain of the gateways.

Fig.~\ref{fig:sourceoflosses}a shows the different sources behind the huge number of uplink transmissions and Fig.~\ref{fig:sourceoflosses}b shows the rate of increase in reference to a network size with 100 devices. The main sources of loss are a) collisions (nrCollisions) b) loss due to channel fading (nrUlost) c) gateway's duty cycle limitation (nrNoDown). nrNoDown metric indicates the number of uplink transmissions of a device that has been received correctly by the gateway but the gateway could not transmit the corresponding downlink (in the two receive windows) due to the duty cycle limitation. In this case, a re-transmission is scheduled. Surprisingly, the collisions are not the main source of loss as it only presents 0.3\% of the losses. This is due to the relatively small network sizes considered in our simulations. For bigger network sizes, the collisions would be a serious source of losses~\cite{abdelfadeel2019free}. The channel fading also has a minimal impact on the losses, about 1.9\%. However, the main source of loss is the duty cycle limitation of the gateway (nrNoDown), which presents about 97.8\% of the losses. This is a very important conclusion that has to be considered when designing the FUOTA routine. More insights are presented in \S\ref{discussion}.

\begin{figure}[!tb]
  \centering
    \includegraphics[width=1\columnwidth]{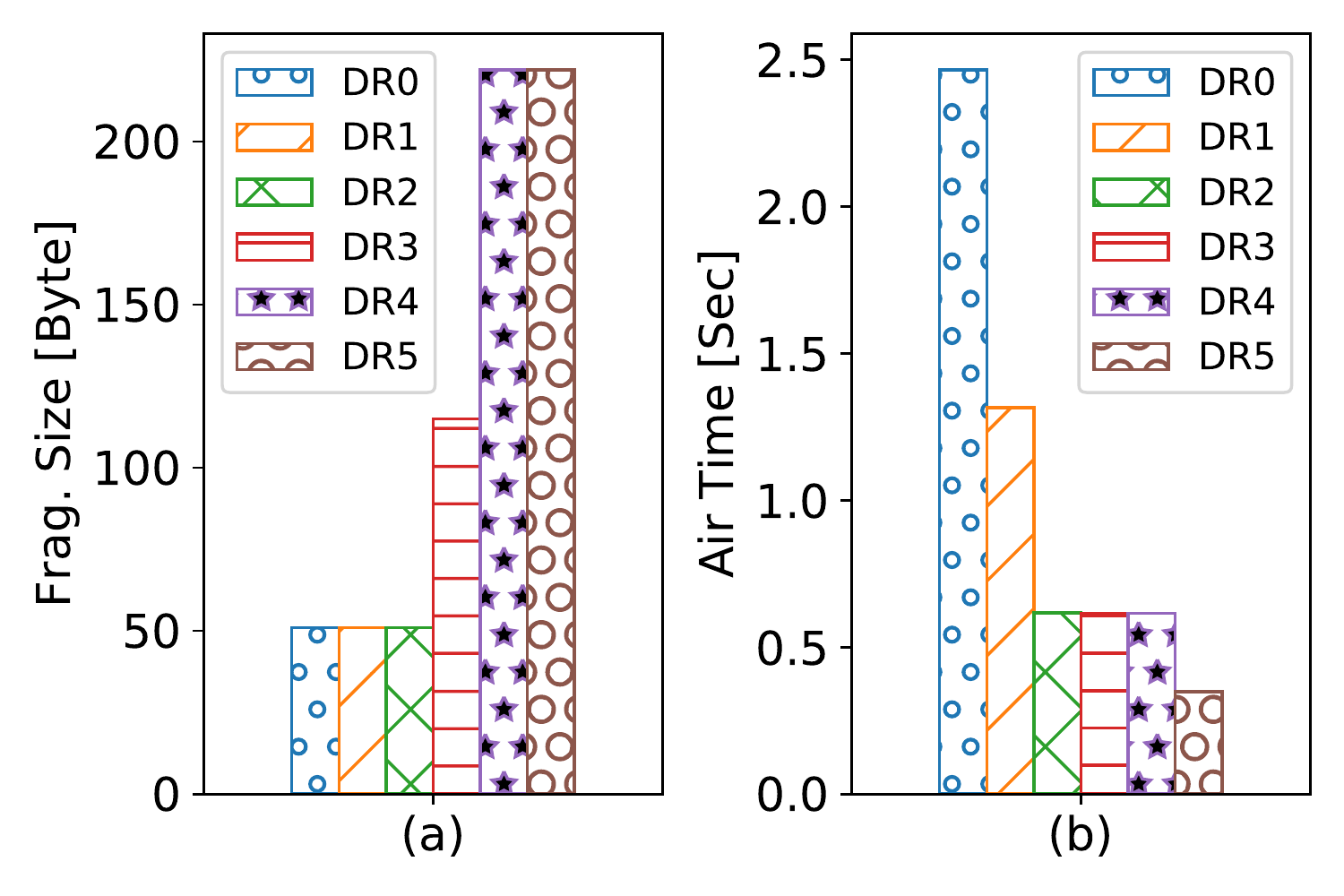}
  \caption{Airtime and size of fragments per data rates}\label{fig:fragsizes_airtime}
  \vspace{-0.35cm}
\end{figure}

\begin{figure}[!tb]
  \centering
    \includegraphics[width=0.8\columnwidth]{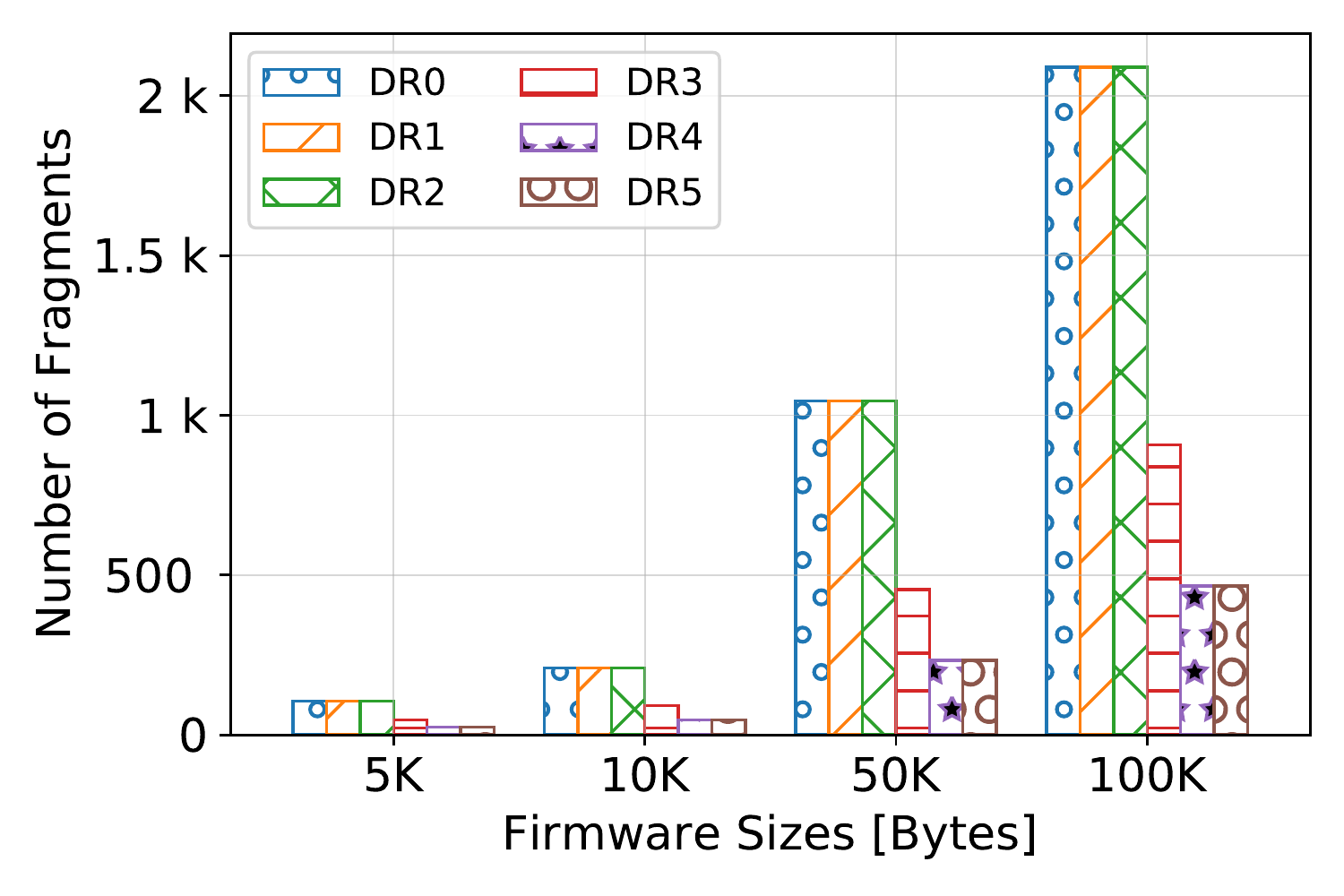}
  \caption{Number of Fragments}\label{fig:nrfragments}
  \vspace{-0.35cm}
\end{figure}

\subsection{Multicast Transmissions Phase Study}

In this subsection, we study the impact on time, energy consumption, and efficiency of the varying configurable parameters for the multicast fragments during the firmware update. These parameters include the data rate used, the class of multicast, either C or B, and the ping periodicity in case of class B. Besides these parameters, we study the impact of the firmware sizes on the aforementioned metrics by considering different sizes: 5kBytes, 10kBytes, 50kBytes, and 100kBytes bytes. As this phase includes only downlink transmissions, the network size does not impact much on the results. Therefore, the presented results are gathered only from a network size with 100 devices, where results can be generalized to the other network sizes.

The fragment size is both directly and inversely proportional to the update time and, thus, the device's energy consumption. On one hand, longer fragment sizes reduce the overall number of fragments, however, the fragments would have a higher probability of error, requiring more redundant fragments. On the other hand, shorter fragment sizes reduce the probability of error but increase the ratio of MAC header (overhead) to payload size, resulting in high overhead. This trade-off has been studied theoretically in~\cite{abdelfadeel2019free} to compute the best packet/fragment size per data rate. Although the calculations have been done for an uplink use case, the conclusion holds true for the downlink as well. The calculations concluded that the impact of packet errors is not as critical as the impact of the MAC header overhead in terms of time and energy consumption. Therefore, long packets for all data rates are better than short packets to reduce the overall number of transmissions and, thus, the impact of MAC headers.

For this reason, the fragment sizes are set to equal the maximum MAC payload sizes (\textit{see} Table.~\ref{tab:lorawan_europe}). Fig.~\ref{fig:fragsizes_airtime} shows the fragment sizes and the airtimes (i.e. transmission times) of one fragment. A clear observation is that the lower the data rate, the higher the airtime even for the same fragment size. For instance, the fragment sizes at DR5 and DR4 are the same but the airtime at DR5 is almost half the airtime at DR4. This is due to the positive relationship between the spreading factor and the airtime~\cite{lora2013calculator}. The fragment sizes also determine the number of fragments (\textit{see} Fig.~\ref{fig:nrfragments}). These numbers along with the airtimes (\textit{see} Fig.~\ref{fig:fragsizes_airtime}b) and the duty cycle limitations affect the firmware update time. Therefore, increasing the data rate, one would expect an increase in the update time and the device's energy consumption. Nevertheless, increasing the data rate, one would also expect an increase in the update efficiency as higher data rates have longer transmission ranges (i.e. lower sensitivity). This trade-off is quantified later in this subsection.

\begin{figure}[!tb]
  \centering
    \includegraphics[width=1\columnwidth]{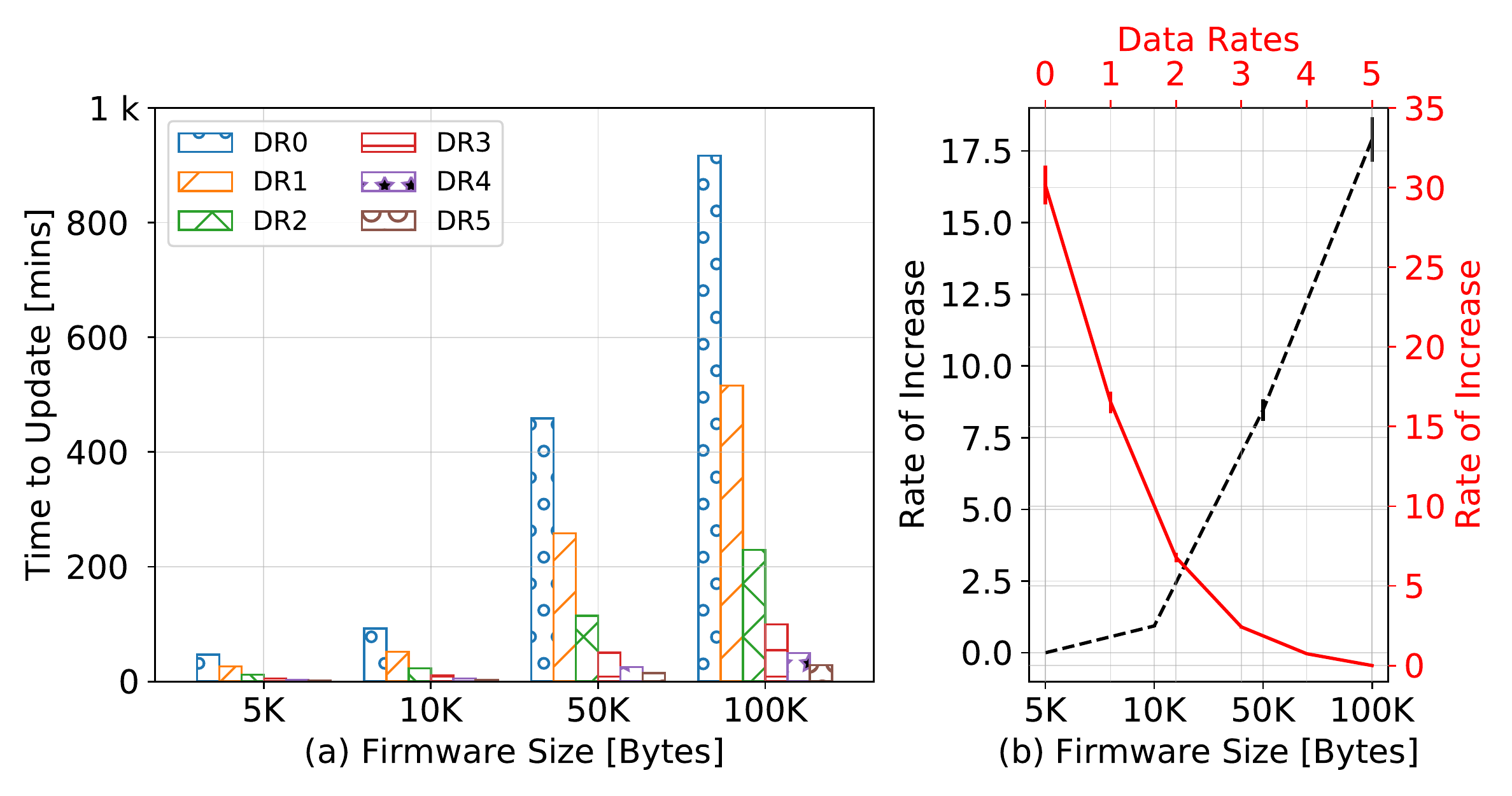}
  \caption{Class C - Time to Update}\label{fig:Ctimetoupdate}
  \vspace{-0.35cm}
\end{figure}

\subsubsection{Multicast Class C}

Fig.~\ref{fig:Ctimetoupdate}a shows the update time across all data rates and for different firmware sizes. Furthermore, Fig.~\ref{fig:Ctimetoupdate}b shows the rate of increase in terms of the data rates and the firmware sizes in reference to DR0 and firmware size of 5kbytes, respectively. The time metric almost doubles with every time the firmware doubles in size. In addition to that, for the same firmware size, the time metric among the data rates shows the same relationship, showing almost 30 times higher when using DR0 than using DR5. This is mainly due to the large number of fragments and the long airtime in the case of DR0, resulting in long silent periods between two successive transmissions due to the duty cycle of the gateway.

\begin{figure}[!tb]
  \centering
    \includegraphics[width=1\columnwidth]{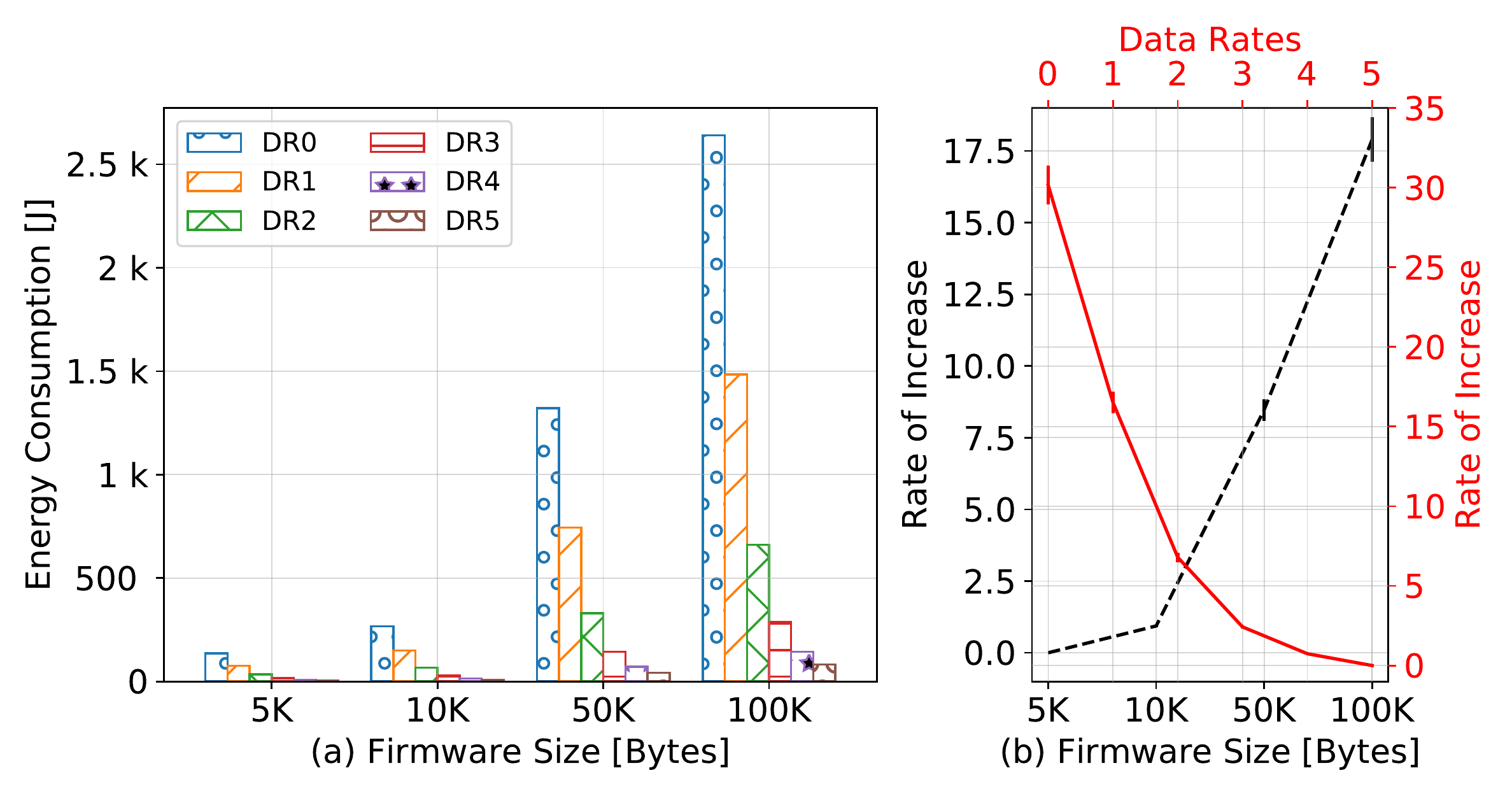}
  \caption{Class C - Energy Consumption}\label{fig:Cenergy}
  \vspace{-0.35cm}
\end{figure}

Fig.~\ref{fig:Cenergy}a shows the device's energy consumption and Fig.~\ref{fig:Ctimetoupdate}b shows the rate of increase in terms of the data rates and the firmware sizes. We can observe that Fig.~\ref{fig:Cenergy}b is almost identical to Fig.~\ref{fig:Ctimetoupdate}b. This is because energy consumption is proportional to the devices' receiving time. In multicast class C, a device is always in a receive mode for the whole time of the update.

\begin{figure}[!tb]
  \centering
    \includegraphics[width=1\columnwidth]{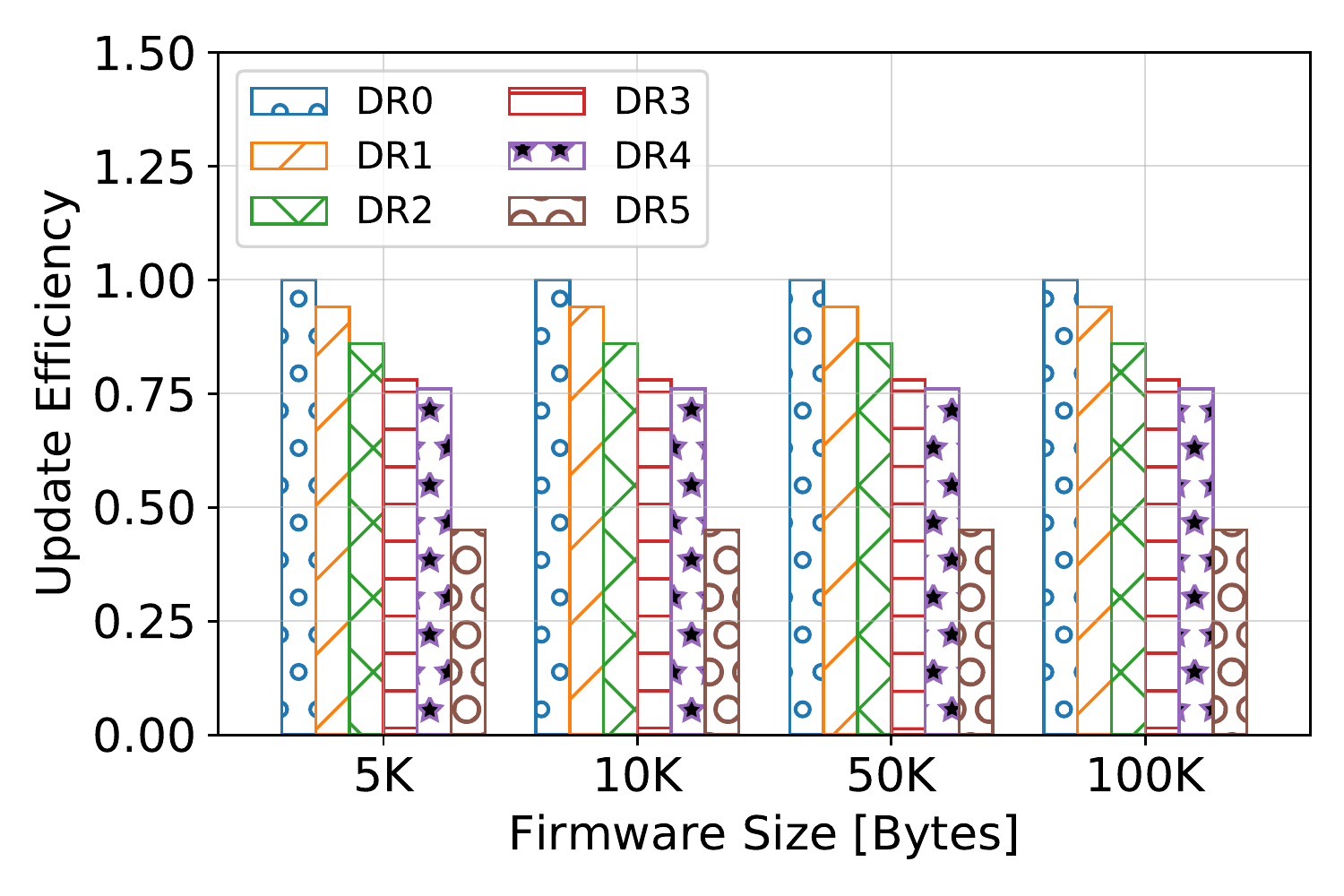}
  \caption{Class C - Update Efficiency}\label{fig:Cupdateefficiency}
  \vspace{-0.35cm}
\end{figure}

From Fig.~\ref{fig:Ctimetoupdate} and Fig.~\ref{fig:Cenergy} it can be observed that for a certain firmware size, the higher the data rate, the lower the update time and the lower the devices' energy consumption. Nevertheless, another factor has to be considered when choosing the data rate, which is the update efficiency. This is because of the fact that the higher the data rate, the shorter the transmission range and, thus, the lower the update efficiency. Fig.~\ref{fig:Cupdateefficiency} shows the update efficiency using all data rates and for different firmware sizes. Using DR0, all devices can be updated at once compared to only 45\% of the devices in the case of using DR5. The update efficiency metric is directly proportional to the considered data rate distribution (\textit{see} Table.~\ref{tab:simparameters}). These results highlight that more than one FUOTA session would be required in the case of using DR5 to update all devices. This would still be acceptable because of the long time and the high energy required in the case of using DR0 (30 times higher than DR5) (\textit{see} Figs.~\ref{fig:Ctimetoupdate} and~\ref{fig:Cenergy}). However, this is subject to the distribution of gateways in the deployment or their mobility, where a gateway may be able to move to reach more devices every time, e.g., a drone based gateway.

\begin{figure}[!tb]
  \centering
    \includegraphics[width=1\columnwidth]{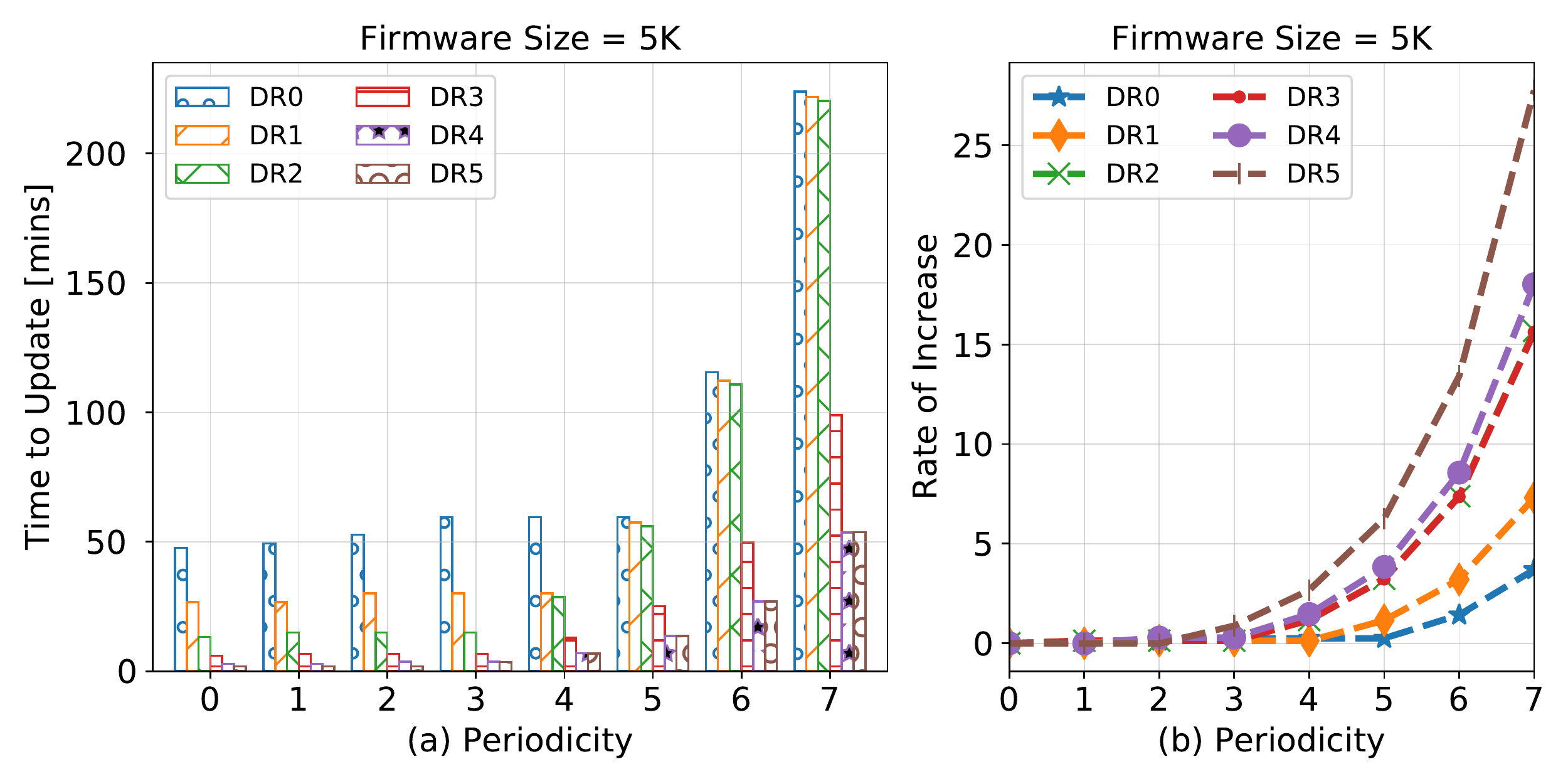}\label{fig:Btimetoupdate}
  \caption{Class B - Time to Update}
  \vspace{-0.35cm}
\end{figure}

\subsubsection{Multicast Class B}

The presented results here are from a firmware image of size 5kbytes only and the results of the other sizes can be roughly estimated using the rate of increase from class C (\textit{see} Fig.~\ref{fig:Ctimetoupdate}b). Fig.~\ref{fig:Btimetoupdate}a shows the update time using all data rates and all ping slot periodicities. Furthermore, Fig.~\ref{fig:Btimetoupdate}b shows the rate of increase of all data rates in reference to ping periodicity $p=0$. As shown, $p=0$ is the best ping periodicity for all data rates as it achieves the lowest update time. This is because of the abundance of ping slots (128 slots) available when $p=0$, which does not limit the downlink transmissions. The results of $p=1$ are close enough to the results of $p=0$. This is because these two ping periodicities are still lower than the duty cycle of the gateway. However, for higher ping periodicities, the time metric increases proportionally to the corresponding ping periodicity. An increase of 17\% is observed in the time metric compared to the results of class C. The reason behind this increase is that the downlink transmissions of class B are performed at the beginning of the ping slots only. In this case, even if the duty cycle of the gateway permits to transmit a new downlink fragment, the transmission has to wait until the beginning of the next ping slot, which prolongs the overall update time.

\begin{figure}[!tb]
  \centering
    \includegraphics[width=1\columnwidth]{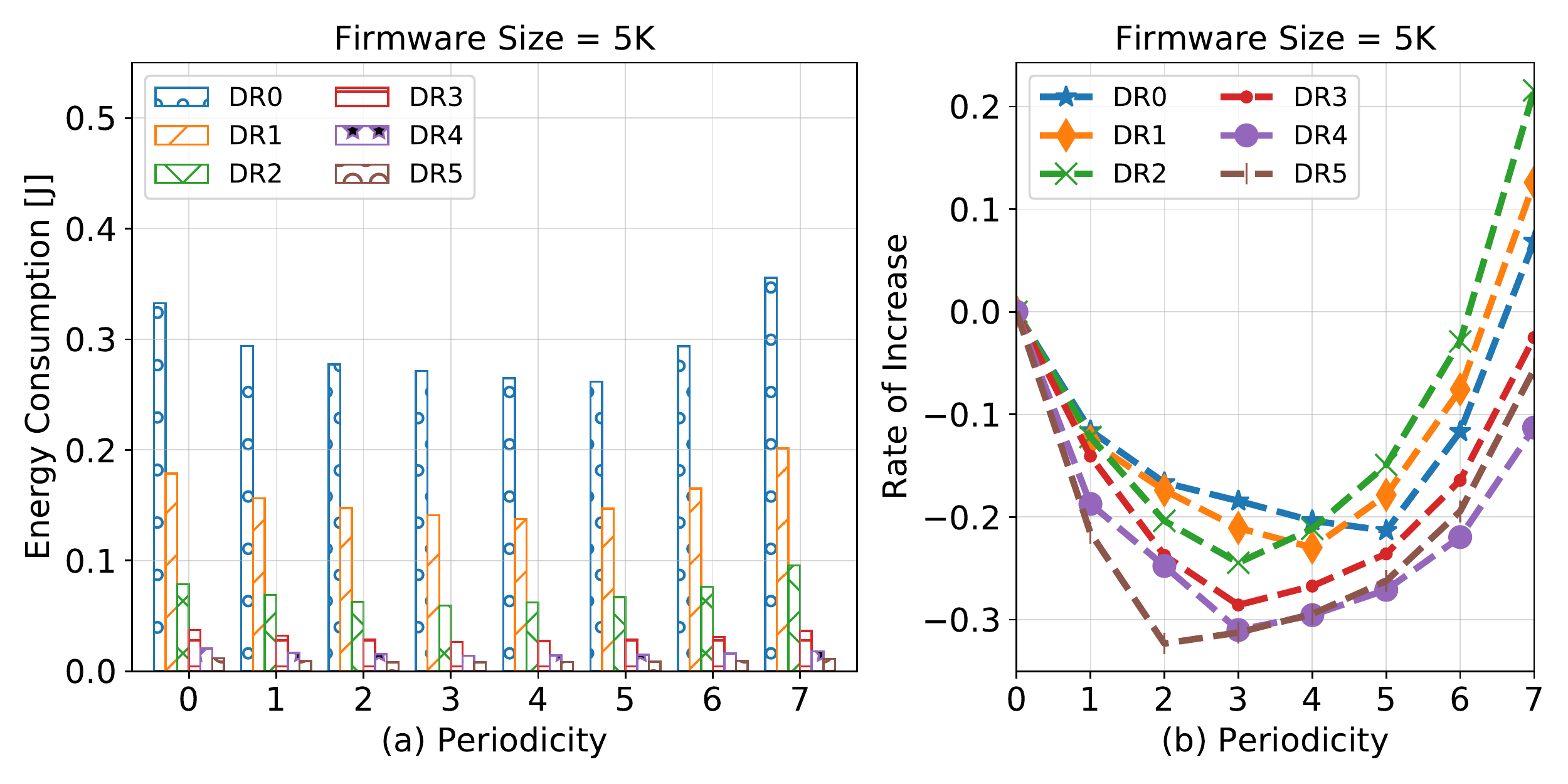}\label{fig:Benergyconsumption}
  \caption{Class B - Energy Consumption}
  \vspace{-0.35cm}
\end{figure}

Fig.~\ref{fig:Benergyconsumption}a shows the device's energy consumption using all data rates and all ping periodicities. The energy consumption is directly proportional to the devices' receiving time. In class B, the radio of a device is in a receive mode only when receiving downlink fragments, receiving gateway beacons, and checking empty ping slots. Otherwise, the radio of a device is in idle mode. The gateway beacons are received to keep the synchronization with the gateway's clock. Checking empty ping slots happens when a ping slot is assigned for the multicast session but the gateway could not transmit in this slot due to the limited duty cycle. In this case, devices stay in a receive mode at the beginning of the empty slots for the time of a packet preamble. Fig.~\ref{fig:Benergyconsumption}b shows the rate of increase of all data rates in terms of the device's energy consumption in reference to ping periodicity $p=0$. It is clear that $p=0$ is not the best periodicity anymore. However, the energy consumption decreases with increasing the periodicity until a certain periodicity, where afterward the energy consumption starts to increase again. This is because of the relationship between the ping slot periodicity and the gateway's duty cycle. If the periodicity is lower than the gateway's duty cycle, devices check a lot of empty ping slots, resulting in high energy consumption. Also, if the periodicity is higher than the gateway's duty cycle, devices have to receive a lot of gateway beacons to keep synchronization, resulting in high energy consumption as well. The best scenario is to have a periodicity close enough to the gateway's duty cycle. As the gateway's duty cycle depends on the data rate used, the best periodicity varies with the data rate. From Fig.~\ref{fig:Benergyconsumption}b, $p=5$ is found to be the best for DR0, $p=4$ for DR1, $p=3$ for DR2, $p=3$ for DR3, $p=3$ for DR4, and $p=2$ for DR5. Considering the best periodicity for each data rate, Class B presents a massive reduction in the device's energy consumption up to 550 times less compared to class C.
 
\section{Discussion} \label{discussion}

Here we discuss some ideas as to how to optimize the FUOTA process for different scenarios. We also provide some further insights based on the above simulation results.

\subsection{Initial Phase}
The initial phase is a prerequisite stage every time a firmware update is to be deployed and, therefore, approaches to reduce its overhead in terms of time and energy consumption are desirable. An efficient approach is particularly desirable for frequent small firmware updates such as security patches whereby the overhead of the initial phase can be much higher than transmitting the firmware update itself. Our simulation results showed that the main source of the overhead is the duty cycle limitation of the gateway. An approach to overcome this might be to use multiple co-located gateways, where the overall overhead can be reduced as the network will have higher duty cycle to handle the downlink transmissions as multiple gateways can be used in parallel. The overhead can be also reduced by minimizing the number of transmissions during the initial phase. This can be achieved by combining multiple commands (e.g., multicast setup and fragmentation setup command) in one command and the gateway can acknowledge both commands with just one packet.

Another overhead reduction mechanism might be to pre-program the devices with the required multicast and fragmentation session information before or during deployment into the field as part of the commissioning stage. In this case, the time synchronization and the multicast start will be the only commands required during the initial phase.

\subsection{Multiple Gateways for Collaborative FUOTA}
The rate at which multicast fragments of the firmware can be transmitted is limited by the duty cycle of the gateway. This limitation leads to prolonged firmware update times, which may not be acceptable for certain applications as the normal device operation is blocked during the update time. Consequently, mechanisms to expedite the multicast transmissions are beneficial. Multiple co-located gateways would be a helpful approach here as well, where the gateways transmit the multicast fragments in a collaborative mechanism. In the case of multicast class C, the gateways can transmit the fragments in a round-robin fashion, taking advantage of those devices that are in a continued receive mode. Consequently, the update time is shortened proportional to the number of the gateways used. For example, using ten gateways in a round-robin fashion would achieve an overall 100\% duty cycle (each gateway transits in the 10\% duty cycle channel).

In the case of multicast class B, a new multicast session is required every time an additional gateway is used when transmitting the fragments. Here, each gateway handles the ping slots of one multicast session and devices wake up to receive all fragments from all gateways. The drawback here is that this adds additional overhead to the initial phase unless these sessions are pre-programmed into the devices before deployment.

\subsection{Network Architecture Planning}
Using a higher data rate to transmit the firmware fragments is always better in terms of time and energy consumption compared to using a lower data rate. However, in the case where firmware updates are transmitted through only one gateway, the simulation results showed that this negatively affects the update efficiency, where lower data rates achieve better coverage and thus better update efficiency than higher data rates. This trade-off can be settled completely towards high data rates in the case of the multiple gateways scenario. If gateways are geographical distributed such that all devices can be reached when using one of the high data rates, we can achieve the shortest update time, the lowest energy consummation, and the maximum update efficiency all at the same time.

However, network architecture planning is usually not done from a network management point of view such as FUOTA but typically from an application point of view in order to achieve required performance metrics, such as high packet delivery ratio. 
Nevertheless, network architecture planning based on the requirements for network management would be beneficial when the rate of FUOTA is expected to be high. An example would be where a LoRaWAN deployment is initially based only on a minimal viable application but over time is upgraded to support a wider range of features. LoRaWAN operators take this approach to expedite their position in the market by deploying their network initially with limited applications and relying on FUOTA to extend and optimize applications over time. In this case, FUOTA sessions could be expected to be scheduled more frequently than usual, e.g., once a month.

The cost of deploying enough gateways in order for all devices to be reached using one of the high data rates could be high, especially for the vast deployments. In this case, using a mobile gateway could be a reasonable approach. In particular, the deployment is divided geographically to small segments, where if the mobile gateway is deployed in the middle of a segment, the devices within that segment can be reached using the desired data rate. Consequently, the gateway moves from one segment to another to update the devices within each segment. This approach is backed by our simulation results that showed gains in terms of time and energy using similar approaches. For example, running approx. 30 FUOTA sessions using the highest data rate is still more beneficial than running one FUOTA using the lowest data rate.

\section{Related Work}

The need for FUOTA has been recognized since the early days of Wireless Sensor Networks (WSNs) whereby the deployment scale and the often remote and inaccessible locations of devices were the main drivers behind this need~\cite{brown2006updating}. Consequently, a lot of research has been carried out, covering different aspects of the firmware update, including protocols for disseminating the update, reducing the size of the update, and executing the update on the devices~\cite{brown2013software}.

The protocols developed for disseminating the firmware update in WSNs are based mainly on the underlying network architecture and protocol stack. In~\cite{hui2004dynamic}, the firmware update is disseminated through multiple paths in multi-hop WSNs. While in~\cite{zimmermann2018split}, the update is targeting network stack modules to reconfigure the network on the fly. The Message Queuing Telemetry Transport (MQTT) protocol is used to disseminate the firmware updates to a fleet of WiFi devices in~\cite{frisch2017over}. For WSNs that enable end-to-end IP connectivity (e.g. 6LoWPAN~\cite{sheng2013survey}), the update can be disseminated using Constrained Application Protocol (CoAP) over UDP~\cite{shelby2014constrained}.

Our work looks at LoRaWAN in contrast to those other network stacks. LoRaWAN does not support the end-to-end IP connectivity and its downlink capability is more limited. In addition to that, LoRaWAN's physical layer supports multiple, quasi-orthogonal data rates and the transmissions are restricted with the duty cycle of the sub-Giga ISM band. These limitations hinder the adaption of legacy protocols for disseminating firmware updates over LoRaWAN. It should be noted that the protocols for FUOTA suggested in this work are network agnostic and, thus, they can be directly adopted in similar networks to LoRaWAN such as Sigfox~\footnote{\url{https://www.sigfox.com/en}}.

\section{Conclusions}

FUOTA is a critical requirement for any long-term deployment of LoRaWAN in order to maintain optimal, safe, and secure operations of the network. In this work, we reviewed new specifications (multicast, fragmentation, and clock synchronization) by LoRa Alliance that make efficient FUOTA possible on top of LoRaWAN. We also evaluated a proposed FUOTA process, showing the impact of the different FUOTA parameters on the performance of the proposed process. The simulation results showed that the initial phase (required for setting up the required sessions) is a bottleneck of the whole process as it is not scalable with the network size. The results also showed that multicast class B achieves 17\% higher update time and 550 times less energy use compared to multicast class C. Additionally, the simulations demonstrated the significant impact of the data rate used on the overall results. For example, DR5 achieves 30\% reduction in update time and device energy consumption compared to DR0. However, this comes at the expense of the firmware update efficiency, where DR5 can only update a portion of the devices (depends on the devices' distribution) every session whilst DR0 can update all devices at once. The ping slot periodicity $p$, in case of multicast class B, impacts the results as well. For all data rates, the fastest firmware update can be done with $p=0$, however, in terms of the energy consumption, the optimal periodicity varies with varying the data rate. In this case, $p=5$ is found to be the best for DR0, $p=4$ for DR1, $p=3$ for DR2, DR3, and DR4, and $p=2$ for DR5.

\section*{Acknowledgment}

This publication has emanated from research conducted with the financial support of Science Foundation Ireland (SFI) and is co-funded under the European Regional Development Fund under Grant Number 13/RC/2077.

\bibliographystyle{IEEEtran}
\bibliography{biblist}

\end{document}